\documentclass{article}

\usepackage{spconfa4, amsmath, amssymb, graphicx, booktabs, multirow, siunitx, float}
\usepackage{xurl}
\usepackage{hyperref}
\hypersetup{colorlinks=true, linkcolor=black, citecolor=black, urlcolor=black}

\usepackage[noadjust]{cite} 


\AtBeginEnvironment{table}{\vspace{-15pt}}
\AtEndEnvironment{table}{\vspace{-17.7pt}}
\AtBeginEnvironment{table*}{\vspace{-6pt}}
\AtEndEnvironment{table*}{\vspace{-18pt}}

\title{Teaching Speech Enhancement Models To Sing:\\ Domain Adaptation from Speech Enhancement to Singing Voice Separation}
\name{Paul A. Bereuter$^{1}$,
      Mark D. Plumbley$^{2}$,     
      Alois Sontacchi$^{1}$
      \thanks{The computing infrastructure used in this work was funded by the digital research infrastructure project ``Interactive Audiovisual Digital Twins of Performance Venues'', supported by the Austrian Federal Ministry of Education, Science and Research (BMFWF). This work was supported by the Engineering and Physical Sciences Research Council (EPSRC) [grant numbers EP/T019751/1, EP/Y028805/1, UKRI397]. For the purpose of open access, the authors have applied a Creative Commons Attribution (CC BY) license to any Author Accepted Manuscript version arising.}
      }
\address{$^{1}$Institute of Electronic Music and Acoustics, University of Music and Performing Arts, Graz, Austria \\
$^{2}$Department of Informatics, King's College London, London, 
United Kingdom}
\begin{document}
\ninept
\maketitle
\begin{abstract}
State-of-the-art speech enhancement models benefit from large-scale labeled datasets, whereas singing voice separation models suffer from limited available training data. To address this limitation, we formulate singing voice separation as domain adaptation from speech enhancement to singing voice separation. We investigate two fine-tuning strategies: full fine-tuning and parameter-efficient fine-tuning using Low-Rank Adaptation (LoRA) on a discriminative and a generative model. Models with either adaptation strategy outperform the same architectures trained from scratch by $0.29\text{-}1.8\text{ dB}$ in Signal-to-Distortion-Ratio. Full fine-tuning yields the highest singing voice separation performance, but catastrophic forgetting degrades speech enhancement performance. LoRA fine-tuning achieves competitive singing voice separation performance while preserving the original speech enhancement capability with only $6\text{-}12\%$ additional parameters compared to the base speech enhancement model. Furthermore, the generative model shows improved generalization to an unseen test set. The results demonstrate that adapting pretrained speech enhancement models is an effective strategy for training singing voice separation models in data-scarce scenarios.
\end{abstract}
\begin{keywords}
singing voice separation, speech enhancement, domain adaptation, low rank adaption
\end{keywords}

\section{Introduction}
\label{sec:intro}
Classical speech enhancement (SE) has primarily addressed acoustic degradations such as environmental noise and reverberation, as tackled in organized challenges such as the Deep Noise Suppression (DNS) Challenge \cite{dns21,dns22,dns23}. The recent transition to more universal enhancement scenarios, such as the Universality, Robustness, and Generalizability for Speech Enhancement (URGENT) Challenge \cite{urgent2025}, reflects a broader shift towards models capable of handling diverse interference conditions beyond traditional denoising tasks. Enabled by increasingly large-scale training datasets covering a wide range of degradations, these developments have led to the emergence of generative and hybrid discriminative–generative enhancement architectures \cite{richter2023speech, lemercier2023storm, scheibler24_interspeech, hirano26refiner}. Broadening the term speech enhancement, these models not only enable separation of speech from diverse interference conditions but also restoration of degraded speech structures. While subsets of large-scale SE datasets include singing voice as part of the target vocal signals, the associated interference conditions typically remain dominated by acoustic distortions such as environmental noise or reverberation, rather than structured musical accompaniment.

In contrast, singing voice separation (SVS) models are commonly trained independently using music datasets such as MUSDB-18-HQ \cite{musdb18} and MoisesDB \cite{moisesdb23}, which provide isolated vocal and accompaniment stems for supervised training under realistic musical interference conditions. However, if combined, these datasets remain limited to approximately 35 hours of labeled data, whereas the recent URGENT challenge corpus \cite{urgent2025} amounts to roughly 700 hours of audio. Discriminative architectures, such as the band-split recurrent neural network (BSRNN) \cite{yu2023bsrnnse, Luo2023mssbsrnn}, as well as score-based generative models (SGM), have nevertheless demonstrated strong performance in both SE and SVS tasks \cite{richter2023speech, bereuter2025gensvs, yunning2026diffvse}, suggesting that similar modeling principles can generalize across these domains. 

We therefore hypothesize that representations learned from large-scale SE pretraining can be transferred to SVS, particularly in data-scarce settings. To investigate this hypothesis, we reformulate SVS as a domain adaptation problem from SE. To this end we explore two adaptation strategies: full fine-tuning of the pretrained SE models and parameter-efficient fine-tuning using LoRA, allowing us to assess the trade-off between increasing SVS performance and preserving the SE capability of the models.
The main contributions of this work are as follows:
\begin{itemize}
    \item We study the transfer of both a discriminative band-split recurrent neural network and a score-based generative SE model to SVS under data-scarce conditions.
    \item We compare full fine-tuning and parameter-efficient fine-tuning using LoRA, and analyze the trade-off between adaptation towards SVS and retaining SE performance.
    \item We analyze out-of-domain performance for the task of singing voice restoration (SVR), to investigate the generalization of the discriminative and generative model variants.
\end{itemize}

\section{Method}
\subsection{Tasks related to separation of vocal signals}
SE and SVS can both be formulated as vocal signal separation problems. In universal SE settings, the observed discrete-time mixture signal $y[n]$ can be written as:
\begin{equation}
y[n] = \mathcal{D}_s\big(s[n]\big) + n[n]
\label{eq:se_noisy}
\end{equation}
where $s[n]$ denotes the clean speech signal, $\mathcal{D}_s(\cdot)$ represents a potentially nonlinear degradation operator affecting the speech signal, and $n[n]$ denotes interfering acoustic signals which may also have been nonlinearly degraded. 
Similarly, singing voice mixtures can be formulated as:
\begin{equation}
y[n] = \mathcal{D}_v\big(v[n]\big) + a[n]
\label{eq:svs_mixture}
\end{equation}
where $v[n]$ denotes the clean singing voice, $\mathcal{D}_v(\cdot)$ represents music production-related processing applied to the vocal signal, and $a[n]$ denotes the accompaniment signal, which itself might have already undergone production-related processing. Depending on the definition of the target signal, two related tasks can be distinguished according to \eqref{eq:svs_mixture}. SVS aims to recover the processed vocal stem $\mathcal{D}_v(v[n])$, while SVR, based on the music source restoration (MSR) framework \cite{yongyi2025MSR}, aims to recover the underlying clean singing voice signal $v[n]$. In this work models are trained exclusively for the SVS objective. We evaluate SVR performance only with an out-of-domain test set to assess generalization. Although, SE, SVS, and SVR share the objective of recovering a target vocal signal from interference and artifacts, they differ in the spectral characteristics of the degradation processes and interference signals. Thus, we formulate SVS as a domain adaptation problem from SE, adapting pretrained SE models to the singing voice domain using limited musical training data. This allows us to investigate whether representations learned from degraded speech transfer to the music production-related processing in the singing domain.
\subsection{Adaptation strategies}\label{sec:adaptation}
To bridge the domain gap between SE and SVS, and leverage the representations learned by SE models, we investigate two adaptation strategies: full fine-tuning and parameter-efficient fine-tuning using LoRA. Both strategies adapt the pretrained SE model for the singing voice domain via fine-tuning on MUSDB18-HQ \cite{musdb18} and MoisesDB \cite{moisesdb23} combined.
\paragraph*{Full fine-tuning\\}
With full fine-tuning all parameters of the pretrained SE model are updated during SVS training. While this approach allows the model to re-learn features which better match the singing voice domain, a performance degradation in the SE domain due to catastrophic forgetting \cite{kirkpatrick2017overcoming} can be expected \cite{lee2020seril}.
\paragraph*{Low-Rank adaptation\\}
Originally proposed for fine-tuning large language models \cite{hu2022lora}, LoRA has been successfully applied to audio foundation models for downstream tasks such as audio classification \cite{Cappellazzo2024LoRAAudioSpecTransformers} and anomalous sound detection \cite{Zheng2024AnomalousSoundDetection}. With LoRA, the pretrained parameters remain fixed, and the model is modified by injecting trainable low-rank matrices into layers carrying out linear transformations. For a frozen pretrained weight matrix $\mathbf{W}_0 \in \mathbb{R}^{d \times k}$, the adapted output $\mathbf{h}$ for an input $\mathbf{x}$ is given by:

\begin{equation}
\mathbf{h} = \mathbf{W}_0 \mathbf{x} + \frac{\alpha}{r} \mathbf{B}\mathbf{A}\mathbf{x}
\label{eq:lora}
\end{equation}
where $\mathbf{B} \in \mathbb{R}^{d \times r}$ and $\mathbf{A} \in \mathbb{R}^{r \times k}$ are trainable matrices with rank $r \ll \min(d, k)$, and $\alpha$ is a scaling factor controlling the contribution of the adaptation branch. During fine-tuning, only $\mathbf{A}$ and $\mathbf{B}$ are updated while the pretrained model weights $\mathbf{W}_0$ remain fixed, so following \eqref{eq:lora}, setting $\alpha = 0$ removes the additive adapter contribution and exactly recovers the original pretrained model.

\begin{table*}[ht]
  \centering
  \caption{Mean and standard deviation for SE (EARS-WHAM), SVS (GenSVS), and SVR (MSRBench) results. The numbers next to ``LoRA'' indicate the rank $r$. Bold numbers indicate the best performance per metric--dataset combination within each model family. For context, a small (S) and large (L) discriminative Mel-RoFormer model (MelRoFo) from~\cite{jensen2024melbandroformer, bereuter2025gensvs} are included. The Noisy row corresponds to unprocessed input mixtures.}
  \label{tab:combined_results}
  \resizebox{0.94\textwidth}{!}{%
  \setlength{\tabcolsep}{3pt}%
  \begin{tabular}{llrrrrrrrrr}
       &    & \multicolumn{3}{c}{\textbf{EARS-WHAM (SE: source domain)}} & \multicolumn{3}{c}{\textbf{GenSVS (SVS: adapted in-domain)}} & \multicolumn{3}{c}{\textbf{MSRBench (SVR: out-of-domain)}} \\
    \cmidrule(lr){3-5} \cmidrule(lr){6-8} \cmidrule(lr){9-11}
       &    & SI-SDR $\uparrow$ & PESQ $\uparrow$ & DistillMOS $\uparrow$ & SDR $\uparrow$ & MR-Loss $\downarrow$ & MERT-MSE $\downarrow$ & SDR $\uparrow$ & MR-Loss $\downarrow$ & MERT-MSE $\downarrow$ \\
    \midrule
    \multirow{5}{*}[0pt]{\rotatebox[origin=c]{90}{\textbf{SGM}}} &   full fine-tuning & \textbf{15.14{\scriptsize $\pm$ 5.63}} & 2.09{\scriptsize $\pm$ 0.52} & 3.53{\scriptsize $\pm$ 0.49} & \textbf{7.62{\scriptsize $\pm$ 6.97}} & 1.283{\scriptsize $\pm$ 0.579} & \textbf{0.112{\scriptsize $\pm$ 0.055}} & \textbf{9.24{\scriptsize $\pm$ 5.62}} & \textbf{1.256{\scriptsize $\pm$ 0.397}} & \textbf{0.104{\scriptsize $\pm$ 0.038}} \\[2pt]
     &   LoRA 16 & 15.05{\scriptsize $\pm$ 5.72} & \textbf{2.49{\scriptsize $\pm$ 0.68}} & \textbf{4.01{\scriptsize $\pm$ 0.54}} & 7.23{\scriptsize $\pm$ 6.76} & \textbf{1.280{\scriptsize $\pm$ 0.561}} & 0.115{\scriptsize $\pm$ 0.056} & 9.05{\scriptsize $\pm$ 5.49} & 1.257{\scriptsize $\pm$ 0.401} & 0.106{\scriptsize $\pm$ 0.040} \\[2pt]
     &   from scratch & 12.97{\scriptsize $\pm$ 6.46} & 1.89{\scriptsize $\pm$ 0.53} & 3.35{\scriptsize $\pm$ 0.60} & 6.94{\scriptsize $\pm$ 6.87} & 1.348{\scriptsize $\pm$ 0.543} & 0.115{\scriptsize $\pm$ 0.054} & 8.82{\scriptsize $\pm$ 5.33} & 1.285{\scriptsize $\pm$ 0.403} & 0.109{\scriptsize $\pm$ 0.040} \\[2pt]
     &   base & 15.05{\scriptsize $\pm$ 5.72} & \textbf{2.49{\scriptsize $\pm$ 0.68}} & \textbf{4.01{\scriptsize $\pm$ 0.54}} & 1.12{\scriptsize $\pm$ 8.27} & 2.089{\scriptsize $\pm$ 1.009} & 0.173{\scriptsize $\pm$ 0.071} & 5.98{\scriptsize $\pm$ 7.91} & 1.562{\scriptsize $\pm$ 0.702} & 0.126{\scriptsize $\pm$ 0.056} \\[2pt]
    \midrule
    \multirow{7}{*}[3pt]{\rotatebox[origin=c]{90}{\textbf{BSRNN}}} &   full fine-tuning & 16.19{\scriptsize $\pm$ 5.68} & 2.52{\scriptsize $\pm$ 0.62} & 3.34{\scriptsize $\pm$ 0.56} & \textbf{9.87{\scriptsize $\pm$ 4.73}} & \textbf{1.148{\scriptsize $\pm$ 0.396}} & \textbf{0.101{\scriptsize $\pm$ 0.047}} & \textbf{10.11{\scriptsize $\pm$ 5.14}} & \textbf{1.125{\scriptsize $\pm$ 0.365}} & \textbf{0.097{\scriptsize $\pm$ 0.035}} \\[2pt]
     &   LoRA 128 & \textbf{17.42{\scriptsize $\pm$ 5.78}} & \textbf{2.80{\scriptsize $\pm$ 0.65}} & \textbf{3.85{\scriptsize $\pm$ 0.67}} & 9.23{\scriptsize $\pm$ 4.49} & 1.197{\scriptsize $\pm$ 0.391} & 0.107{\scriptsize $\pm$ 0.047} & 9.76{\scriptsize $\pm$ 5.07} & 1.188{\scriptsize $\pm$ 0.392} & 0.104{\scriptsize $\pm$ 0.038} \\[2pt]
     &   LoRA 32 & \textbf{17.42{\scriptsize $\pm$ 5.78}} & \textbf{2.80{\scriptsize $\pm$ 0.65}} & \textbf{3.85{\scriptsize $\pm$ 0.67}} & 8.92{\scriptsize $\pm$ 4.56} & 1.205{\scriptsize $\pm$ 0.425} & 0.109{\scriptsize $\pm$ 0.048} & 9.56{\scriptsize $\pm$ 5.10} & 1.210{\scriptsize $\pm$ 0.406} & 0.108{\scriptsize $\pm$ 0.041} \\[2pt]
     &   LoRA 16 & \textbf{17.42{\scriptsize $\pm$ 5.78}} & \textbf{2.80{\scriptsize $\pm$ 0.65}} & \textbf{3.85{\scriptsize $\pm$ 0.67}} & 8.76{\scriptsize $\pm$ 4.61} & 1.247{\scriptsize $\pm$ 0.485} & 0.111{\scriptsize $\pm$ 0.050} & 9.51{\scriptsize $\pm$ 5.15} & 1.251{\scriptsize $\pm$ 0.445} & 0.110{\scriptsize $\pm$ 0.042} \\[2pt]
     &   from scratch & 11.78{\scriptsize $\pm$ 6.28} & 1.72{\scriptsize $\pm$ 0.41} & 2.97{\scriptsize $\pm$ 0.45} & 8.05{\scriptsize $\pm$ 4.35} & 1.256{\scriptsize $\pm$ 0.496} & 0.125{\scriptsize $\pm$ 0.046} & 9.19{\scriptsize $\pm$ 4.93} & 1.257{\scriptsize $\pm$ 0.406} & 0.119{\scriptsize $\pm$ 0.041} \\[2pt]
     &   base & \textbf{17.42{\scriptsize $\pm$ 5.78}} & \textbf{2.80{\scriptsize $\pm$ 0.65}} & \textbf{3.85{\scriptsize $\pm$ 0.67}} & 3.48{\scriptsize $\pm$ 9.47} & 1.815{\scriptsize $\pm$ 0.923} & 0.165{\scriptsize $\pm$ 0.074} & 7.15{\scriptsize $\pm$ 6.47} & 1.415{\scriptsize $\pm$ 0.595} & 0.124{\scriptsize $\pm$ 0.054} \\[2pt]
    \midrule
    \multirow{2}{*}[2.5pt]{\rotatebox[origin=c]{90}{\shortstack{\textbf{Ref.} \\ \cite{jensen2024melbandroformer, bereuter2025gensvs}}}} &   MelRoFo (S) & 13.46{\scriptsize $\pm$ 6.71} & \textbf{1.91{\scriptsize $\pm$ 0.56}} & 2.90{\scriptsize $\pm$ 0.57} & 8.98{\scriptsize $\pm$ 4.70} & 1.393{\scriptsize $\pm$ 0.568} & 0.110{\scriptsize $\pm$ 0.049} & 10.47{\scriptsize $\pm$ 5.16} & 1.131{\scriptsize $\pm$ 0.389} & 0.094{\scriptsize $\pm$ 0.035} \\[2pt]
     &   MelRoFo (L) & \textbf{14.37{\scriptsize $\pm$ 6.68}} & 1.90{\scriptsize $\pm$ 0.58} & \textbf{3.13{\scriptsize $\pm$ 0.60}} & \textbf{11.78{\scriptsize $\pm$ 4.91}} & \textbf{1.217{\scriptsize $\pm$ 0.419}} & \textbf{0.084{\scriptsize $\pm$ 0.039}} & \textbf{11.93{\scriptsize $\pm$ 5.63}} & \textbf{1.077{\scriptsize $\pm$ 0.377}} & \textbf{0.082{\scriptsize $\pm$ 0.032}} \\[2pt]
    \midrule
     &   Noisy & 6.02{\scriptsize $\pm$ 6.33} & 1.28{\scriptsize $\pm$ 0.27} & 2.76{\scriptsize $\pm$ 0.55} & -4.28{\scriptsize $\pm$ 4.20} & 2.547{\scriptsize $\pm$ 1.343} & 0.199{\scriptsize $\pm$ 0.042} & -1.50{\scriptsize $\pm$ 4.12} & 2.110{\scriptsize $\pm$ 0.887} & 0.188{\scriptsize $\pm$ 0.040} \\
    \bottomrule
  \end{tabular}
  }%
\end{table*}

\subsection{Architectures}
We investigate domain adaptation from SE to SVS for two model architectures, representing both discriminative and generative modeling paradigms. Both models were pretrained on large-scale SE datasets. For the discriminative model we employ the Band-Split Recurrent Neural Network (BSRNN) \cite{yu2023bsrnnse, Luo2023mssbsrnn}. BSRNN divides the complex-valued input STFT into multiple frequency sub-bands, each of which are processed by recurrent neural networks (RNNs) that model both temporal sequences and sub-band dependencies. The sub-band outputs are combined to generate a complex-valued mask that is applied to the noisy input mixture, to estimate the target vocal signal. This mask-based approach provides a direct regression from noisy mixtures to clean vocal signals. For the generative architecture, we employ the score-based generative model for speech enhancement (SGMSE) \cite{richter2023speech}, a score-based diffusion model operating on complex-valued STFT representations. The model learns to iteratively estimate clean vocal signals from noise corrupted mixtures through a diffusion process governed by stochastic differential equations (SDEs) \cite{song2021scorebased}. Unlike the deterministic BSRNN, this stochastic framework enables the generation of multiple realizations of clean vocal estimate from a single input mixture.

\section{Experimental Setup}
\subsection{Pretraining and adaptation}
All models in this work operate at a sampling rate of \SI{48}{\kilo\hertz} and are initialized from pretrained checkpoints. For the BSRNN, weights pretrained on the URGENT dataset (approx. \SI{700}{\hour} of audio) from \cite{urgent2025} are employed. The SGM, which utilizes a noise-conditional score network (NCSN++) \cite{song2021scorebased} as its backbone, was pretrained for SE on the EARS-WHAM dataset \cite{richter2024ears} (approx. \SI{87}{\hour}). 

\subsection{Datasets}\label{sec:datasets}
To adapt the models for the singing voice domain, we utilize two labeled multi-stem datasets: MUSDB18-HQ \cite{musdb18} and MoisesDB \cite{moisesdb23}. These datasets provide isolated vocal and accompaniment stems, enabling supervised training with musical interferences. Together, they comprise approximately \SI{35}{\hour} of audio. To increase training data variability, random gains and random mixing are applied as data augmentation. SVS evaluation is performed on the \SI{5}{\second} excerpts of the MUSDB18-HQ test set from \cite{bereuter2025gensvs} (GenSVS), SVR evaluation on the MSRBench test set \cite{yongyi2025MSRBench}, and SE evaluation on the first \SI{5}{\second} of the EARS-WHAM test set \cite{richter2024ears}. All MUSDB18-HQ audio has been upsampled from \SI{44.1}{\kilo\hertz} to \SI{48}{\kilo\hertz} to match the operating sampling rate of the models. The mixtures and targets of the MSRBench test set exhibit gain mismatches, leading to mismatched gains between predictions and targets. So prior to metric calculation we match the loudness of the singing voice predictions with the respective target levels according to the EBU R128 standard \cite{EBU-R128}.
\subsection{Model configurations}
The BSRNN architecture published with \cite{urgent2025} employs an STFT encoder/decoder with a Hann window, an FFT size of $N_{\text{fft}} = 960$, and a hop size of $480$ samples. The band-split module decomposes the input into 34 non-overlapping subbands ($20\times\SI{200}{\hertz}$ bands for $0-\SI{4}{\kilo\hertz}$, $6\times\SI{500}{\hertz}$ for $4-\SI{7}{\kilo\hertz}$, $7\times\SI{2}{\kilo\hertz}$ for $7-\SI{21}{\kilo\hertz}$ and $1\times\SI{3}{\kilo\hertz}$ for $21-\SI{24}{\kilo\hertz}$), each of which is projected to a common feature dimension corresponding to 196 hidden channels per band. The features are stacked and processed by a band and sequence modeling module consisting of six recurrent layers. Each layer applies two bidirectional Long Short-Term Memory (BiLSTM) blocks sequentially: a sequence-level BiLSTM that models temporal dependencies within each band, followed by a band-level BiLSTM that captures inter-band dependencies across subbands\cite{Luo2023mssbsrnn}. Following \cite{urgent2025}, all BSRNN models are trained with a weighted composite loss consisting of \SI{20}{\percent} L1-waveform and \SI{80}{\percent} scale-invariant Signal-to-Distortion-Ratio (SI-SDR) loss.

The pretrained checkpoint of the SGM model from \cite{richter2024ears} operates on spectrograms with $N_{\text{fft}} = 1534$ and a hop size of $384$ samples. The diffusion process is defined by an Ornstein-Uhlenbeck Variance-Exploding (OUVE) SDE \cite{richter2023speech}, which describes how clean vocal signals are gradually corrupted into noisy mixtures and, in reverse, how they are recovered during inference. We use the same noise schedule parameters as the ones for the base SE model \cite{richter2024ears}. All SGM models are trained with the conditional denoising score-matching objective from \cite{richter2023speech}. For inference we use a Predictor-Corrector (PC) sampler with an Annealed Langevin Dynamics (ALD) corrector \cite{song2019ald}, utilizing $N = 45$ diffusion steps, a step size of $0.5$ and $2$ correction steps, the same as in~\cite{bereuter2025gensvs}. To ensure statistical robustness in the evaluation of the generative model, we generate 10 realizations of the clean vocal estimate for each mixture audio sample within the three test sets. Realization-level evaluation metrics are averaged per audio sample before computing the results statistics presented in \autoref{sec:results}.
\subsection{LoRA adaptation parameters}
When applying LoRA to the BSRNN model, the rank is treated as a hyperparameter. We evaluate ranks $r \in \{16, 32, 128\}$, $r=8$ was tested but its separation performance was insufficient. The selected ranks increase model size (see \autoref{tab:model_parameters}) while remaining computationally feasible. The scaling factor is chosen heuristically as $\alpha=2r$ for all ranks. LoRA is applied to linear and convolutional layers of the BSRNN, initialized using the default strategy of \texttt{peft} \cite{peft}, where the LoRA matrices in \eqref{eq:lora} are set to $\mathbf{B} = \mathbf{0}$ and $\mathbf{A}$ follows a Kaiming uniform distribution \cite{He2015Kaiming}. Bias adaptation is disabled and a dropout rate of 0.005 was used throughout. For adaptation of the SGM, we fix $r=16$ and $\alpha=32$, as a single training run requires approximately six days on three NVIDIA RTX 6000 Blackwell Max-Q GPUs, making systematic rank search infeasible. Here LoRA is applied to all ResNet time-embedding linear projections and ResNet convolutions of the NCSN++ backbone. SGM models are trained with a batch size of 4 for full fine-tuning and 3 for LoRA, using learning rates of $1\times10^{-4}$ and $2\times10^{-4}$, respectively. All fine-tuned BSRNN models use a batch size of 14 and a learning rate of $5\times10^{-5}$. For from scratch training, both model families use a learning rate of $1\times10^{-4}$, with batch sizes of 4 for SGM and 14 for BSRNN. All models are trained for 550 epochs. Checkpoints with the highest Signal-to-Distortion-Ratio (SDR) on the GenSVS test set are selected for testing.
\subsection{Evaluation metrics}\label{sec:metrics}
To assess SVS and SVR performances, we use the SDR \cite{vincent2006} computed with \cite{torchmetrics22}, employing a filter-based decomposition that makes it more suitable than the SI-SDR for signals subject to music production-related processing. Additionally, we employ embedding-based and perceptually motivated metrics: a mean squared error in the embedding space of the Acoustic Music Understanding Model with Large-Scale Self-supervised Training (MERT-MSE) \cite{li2024mert}, computed using the \texttt{gensvs} package \cite{bereuter2025gensvs}, and a perceptually A-weighted multi-resolution loss (MR-Loss) implemented with \texttt{auraloss} \cite{steinmetz2020auraloss}. These metrics have been shown to correlate well with perceptual audio quality ratings from listening tests \cite{bereuter2025gensvs, bereuter2026daga}. SVR performance on the MSRBench test set is evaluated using loudness-matched predictions and targets to avoid gain-related metric bias (see \autoref{sec:datasets}). For SE, we report reference-based metrics including SI-SDR \cite{sisdrRoux2019} and Perceptual Evaluation of Speech Quality (PESQ) \cite{rix2001pesq}. In addition, we include DistillMOS \cite{stahl2025}, a referenceless metric that predicts mean opinion score (MOS)-like speech quality ratings. All SE metrics but DistillMOS \cite{stahl2025} are computed using \texttt{torchmetrics} \cite{torchmetrics22}. 
\section{Results and Discussion}
\begin{figure}[t]
  \centering
  \centerline{\includegraphics[width=0.6\columnwidth]{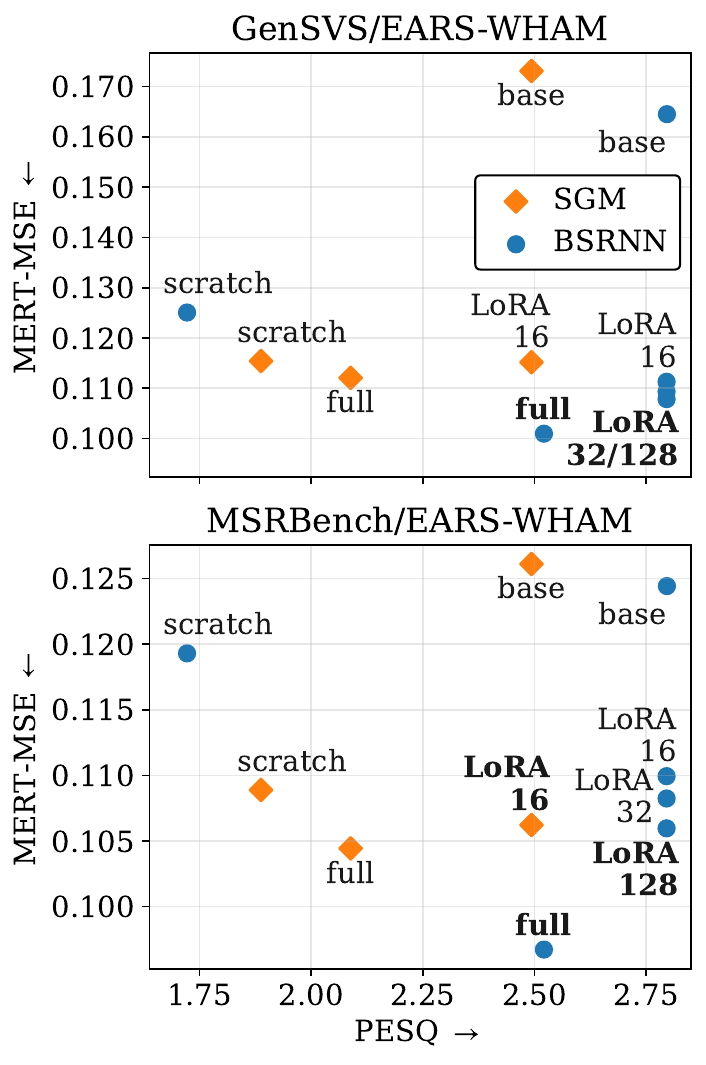}}
  \caption{Trade-off between MERT-MSE and PESQ for SE on the EARS-WHAM test set. MERT-MSE was evaluated on the GenSVS (top) and on the MSRBench test set (bottom). Models in the lower right corner indicate a better trade-off. \vspace*{-5mm}}
  \label{fig:mertmse_vs_pesq}
\end{figure}
\paragraph*{Performance trade-off in domain adaptation\\}
As shown in \autoref{tab:combined_results} and \autoref{fig:mertmse_vs_pesq}, full fine-tuning achieves the highest scores across both singing voice datasets. On GenSVS (in-domain), BSRNN and SGM reach an SDR of \SI{9.87}{\decibel} and \SI{7.62}{\decibel}, and on MSRBench (out-of-domain) \SI{10.11}{\decibel} and \SI{9.24}{\decibel}, respectively. However, this comes at the cost of a degraded SE performance with PESQ dropping by $0.28$ for BSRNN and $0.40$ for SGM, revealing a trade-off between target-domain performance and source-domain preservation. With LoRA fine-tuning the original SE performance can be preserved, as the additive LoRA contribution  can be disabled (see \autoref{sec:adaptation}). At the same time, LoRA achieves competitive SVS and SVR performance. In the trade-off plot in \autoref{fig:mertmse_vs_pesq}, LoRA variants cluster around a PESQ of $2.5-2.75$ while achieving lower MERT-MSE values than the base models. For example, the LoRA fine-tuned BSRNN (rank 32) attains an SDR of \SI{8.92}{\decibel} on GenSVS and \SI{9.56}{\decibel} on MSRBench, retaining over \SI{90}{\percent} of full fine-tuning performance. This is achieved with only \SI{12.12}{\percent} of additional parameters (see \autoref{tab:model_parameters}) compared to the base model. Increasing the rank from $r=16$ to $r=128$ yields only marginal gains (\SI{0.47}{\decibel} on GenSVS and \SI{0.25}{\decibel} on MSRBench), while the parameter count rises to \SI{48.49}{\percent} above the base model, indicating that lower ranks suffice for this domain shift. Both reference Mel-RoFormer (MelRoFo)~\cite{wang_melroformer_2024} models are added for context: while both are trained from scratch for SVS, our adapted models match or, in some cases, surpass MelRoFo (S)~\cite{bereuter2025gensvs} (parameter count comparable to SGM), whereas MelRoFo (L)~\cite{jensen2024melbandroformer} achieves the best SVS and SVR scores at more than three times the parameters and with additional undisclosed training data. We acknowledge that achieving state-of-the-art performance is not the primary focus of this work, which instead aims at highlighting the benefits of SE pretraining.

\label{sec:results}


\begin{table}[!ht]
  \centering
  \caption{Mean performance delta between GenSVS and MSRBench. Bold values denote the best improvement per row and metric.}
  \label{tab:delta_results}
  \resizebox{0.8\columnwidth}{!}{%
  \setlength{\tabcolsep}{4pt}%
  \begin{tabular}{lrrrr}
       & \multicolumn{2}{c}{\textbf{SGM}} & \multicolumn{2}{c}{\textbf{BSRNN}} \\
    \cmidrule(lr){2-3} \cmidrule(lr){4-5}
    Model variant & $\Delta$ SDR & $\Delta$ MSE & $\Delta$ SDR & $\Delta$ MSE \\
    \midrule
    full fine-tuning & \textbf{1.62} & \textbf{-0.008} & 0.24 & -0.004 \\[5pt]
    LoRA 16 & \textbf{1.82} & \textbf{-0.009} & 0.74 & -0.001 \\[5pt]
    from scratch & \textbf{1.89} & \textbf{-0.007} & 1.15 & -0.006 \\[5pt]
    \bottomrule
  \end{tabular}
  }%
\end{table}
\paragraph*{Comparative analysis and generalization\\}
The results in \autoref{tab:combined_results} and \ref{tab:delta_results} highlight the robust out-of-domain generalization of score-based generative models. The clearest evidence comes from the ``from scratch'' configuration, where both architectures are trained exclusively on the same SVS dataset (see \autoref{sec:datasets}). Here, SGM shows more improvement from GenSVS to MSRBench in both $\Delta$SDR and $\Delta$MERT-MSE (see \autoref{tab:delta_results}), while achieving lower absolute MERT-MSE on MSRBench than the ``from scratch'' BSRNN. This trend persists for ``LoRA'' configurations, where SGM again achieves lower MERT-MSE on MSRBench despite using a substantially smaller pretraining dataset (\SI{87}{\hour} vs. \SI{700}{\hour}). The fully fine-tuned variants demonstrate a similar trend. In absolute terms, BSRNN achieves a lower MERT-MSE on MSRBench, compared to SGM (0.097 vs. 0.104). This result is, however, undermined by their differing pretraining datasets.

\begin{table}[ht]
  \centering
  \caption{Parameter counts and complexity: total parameters (M), relative parameter increase (\%) and GMACs/s measured with \cite{ptflops}.}
  \label{tab:model_parameters}
  \resizebox{0.7\columnwidth}{!}{%
  \setlength{\tabcolsep}{4pt}%
  \begin{tabular}{llrrr}
    & Model & \shortstack[c]{Total\\Params\\{[M]}} & \shortstack[c]{Add.\\Params\\{[\%]}} & \shortstack[c]{Complexity\\{[GMACs/s]}} \\
    \midrule
    \multirow{4}{*}{\rotatebox[origin=c]{90}{\textbf{SGM}}} & full fine-tuning & 64.74 & 0.00 & 399.64 \\
     & LoRA 16 & 69.76 & 7.75 & 895.07 \\
     & from scratch & 64.74 & 0.00 & 399.64 \\
     & base & 64.74 & 0.00 & 399.64 \\
    \midrule
    \multirow{6}{*}{\rotatebox[origin=c]{90}{\textbf{BSRNN}}} & full fine-tuning & 37.80 & 0.00 & 84.31 \\
     & LoRA 128 & 56.13 & 48.49 & 91.18 \\
     & LoRA 32 & 42.38 & 12.12 & 86.03 \\
     & LoRA 16 & 40.09 & 6.06 & 85.17 \\
     & from scratch & 37.80 & 0.00 & 84.31 \\
     & base & 37.80 & 0.00 & 84.31 \\
    \bottomrule
  \end{tabular}
  }%
\end{table}
\section{Conclusion}
\label{sec:conclusion}
In this work, we investigate the adaptation of pretrained speech enhancement (SE) models to singing voice separation (SVS). While full fine-tuning achieves the best SVS performance, our results show that LoRA enables domain adaptation while preserving the original model capabilities. Leveraging its additive structure, LoRA retains the original SE performance while achieving competitive SVS results, thereby mitigating catastrophic forgetting and enabling efficient repurposing of the used pretrained models. This task-specific adaptation makes LoRA ideal for universal source separation frameworks targeting multiple audio domains, e.g. speech and singing. 

Comparing discriminative (BSRNN) and generative (SGM) models, we further find that the generative model generalizes better to out-of-domain data in the singing voice restoration task, particularly in terms of MERT-MSE across both from-scratch training and LoRA-based adaptation. Despite being pretrained on a substantially smaller dataset and using lower adaptation ranks, the generative approach shows a larger performance improvement from the in-domain (GenSVS) to the out-of-domain (MSRBench) test set. Overall, our results highlight that SE pretraining transfers effectively to SVS and that lightweight LoRA adapters allow for efficient domain adaptation while preserving source-domain performance. 

Future work will explore pretraining and fine-tuning strategies across audio sources, including identifying suitable source domains for transfer to instrument-specific separation and extending to multi-stem music separation.
The source code is openly available\footnote{\url{https://github.com/pablebe/se2svs}} and audio examples are provided on the companion website\footnote{\url{https://pablebe.github.io/se2svs-webpage/}}.

\clearpage
\bibliographystyle{IEEEtran}
\bibliography{refs}

\end{document}